\documentclass[reqno,dvips,12pt]{article}

\usepackage{epsfig,graphicx,amsmath,amssymb,amsfonts}
\newcommand{\dt}{\delta}

\newcommand{\ov}{\overline}
\newcommand{\wt}{\widetilde}

\newcommand{\nn}{\nonumber}

\newcommand{\bR}{{\bf R}}

\newcommand{\wh}{\widehat}
\newcommand{\str}{\stackrel}
\newcommand{\pa}{\partial}

\newcommand{\be}{\beta}
\newcommand{\la}{\lambda}
\newcommand{\al}{\alpha}

\newcommand{\Om}{\Omega}
\newcommand{\R}{\mathbb R}
\newcommand{\La}{\Lambda}

\newcommand{\De}{\Delta}
\newcommand{\op}{\operatorname}
\newcommand{\si}{\sigma}
\renewcommand{\r}{\right)}
\renewcommand{\l}{\left(}

\begin{document}

\title {Uniform Asymptotics in the Problem of Superfluidity\\
of Classical Liquids in Nanotubes}
\author{V.~P.~Maslov}
\date{}
\maketitle

\begin{abstract}
In the preceding papers (see \cite{Arxiv_1, ArXiv_2}),
the superfluidity of the classical liquid was proved
under the assumption that the parameters~$N$ and~$r$,
where~$N$ is the particle number and~$r$ it the capillary radius,
tend respectively to infinity and to zero
so that $\frac 1N \ll \frac rR$, where $R$ is the capillary length.
In the present paper, this assumption is removed.
\end{abstract}

\textbf{1.} We first note that solutions of the variational
equation for the Vlasov equation do not coincide with the
classical limit for the variational equations corresponding to
the mean-field equations in quantum theory. We consider
mean-field equations of the form

\begin{equation}\label{Hart}
ih\frac{\pa}{\pa t}\varphi^t (x)= \bigg(-\frac{h^2}{2m}\Delta
+W_t(x)\bigg) \varphi^t(x), \qquad W_t(x)=U(x)+\int
V(x,y)|\varphi^t(y)|^2dy,
\end{equation}
with the initial condition  $\varphi|_{t=0}=\varphi_0$, where
$\varphi_0 \in W_2^\infty (\bR^\nu), \int dx|\varphi_0(x)|^2=1$.

To find an asymptotic representation of the complex-germ type
\cite{VKB}, we consider a system consisting of the Hartree
equation and its conjugate equation. We then take the system of
variational equations for it and replace the variations
$\delta\varphi$  and $\delta\varphi^*$ with the independent
functions $F$ and $G$. For $F$ and $G$, we obtain the system of
equations:
\begin{eqnarray}\label{Shv1}
&& \ i\frac{\pa F^t(x)}{\pa t} = \int dy
\bigg(\frac{\delta^2H}{\delta\varphi^*(x)\delta\varphi(y)}F^t(y)+
\frac{\delta^2H}{\delta\varphi^*(x)\delta^*\varphi(y)}G^t(y)\bigg);
\\
&&-i\frac{\pa G^t(x)}{\pa t} = \int dy
\bigg(\frac{\delta^2H}{\delta\varphi(x)\delta\varphi(y)}F^t(y)+
\frac{\delta^2H}{\delta\varphi(x)\delta^*\varphi(y)}G^t(y)\bigg).
\nn
\end{eqnarray}
Roughly speaking, the classical equations can be obtained from
the quantum ones using a substitution of the form $\varphi= \chi
e^{\frac ih S}$ (the WKB method), $\varphi^*= \chi^* e^{\frac ih
S^*}, \ S=S^*, \ \chi=\chi(x,t)\in C^\infty, \ S=S(x,t)\in
C^\infty$.

For variational equations, it is natural to vary not only the
limit equation for $\chi$ and $\chi^*$ but also the functions $S$
and $S^*$. This gives an important new term in the solution of the
equation for collective oscillations. We consider this fact for
the simplest example investigated in the famous work by
Bogoliubov on a weakly nonideal Bose gas \cite{Bogol}.

Let $U=0$ for ~(\ref{Hart}) in a three-dimensional box with edge
length $L$, and let the $L$-periodicity condition be imposed on
the wave functions in this case (i.e., a problem on the torus with
generators $L, L$, and $L$ is considered). Then the function

\begin{equation}\label{bog1}
\varphi(x)=L^{-3/2} e^{i/h(px-\Omega t)},
\end{equation}
with $p=2\pi n/L$, where $n$ is an integer-valued vector,
satisfies  ~(\ref{Hart}) with
\begin{equation}\label{bog2}
\Omega=\frac{p^2}{2m}+L^{-3}\int dx V(x).
\end{equation}

We consider functions $F^{(\lambda)}(x)$ and $G^{(\lambda)}(x)$,
where $\lambda=2\pi {n}/L, {n}\neq 0$ of the form
\begin{eqnarray}\label{bog3}
&& F^{(\lambda)t}(x)= L^{-3/2}\rho_\lambda e^{\frac
ih|(p+\lambda)x +(\beta-\Omega)t|}, \nn \\
&& G^{(\lambda)t}(x)= L^{-3/2}\sigma_\lambda e^{\frac
ih|(-p+\lambda)x +(\beta+\Omega)t|},
\end{eqnarray}
where
\begin{eqnarray}\label{bog4}
&&-\beta_\lambda\rho_\lambda=\bigg(\frac{(p+\lambda)^2}{2m}-\frac{p^2}{2m}
+\wt{V}_\lambda\bigg)\rho_\lambda+\wt{V}_\lambda\sigma_\lambda, \nn \\
&&\beta_\lambda\rho_\lambda=\bigg(\frac{(p-\lambda)^2}{2m}-\frac{p^2}{2m}
+\wt{V}_\lambda\bigg)\sigma_\lambda+\wt{V}_\lambda\rho_\lambda,\\
&&|\sigma_\lambda|^2 -|\rho_\lambda|^2=1, \qquad
\wt{V}_\lambda=L^{-3}\int dx V(x)e^{\frac ih \lambda x}. \nn
\end{eqnarray}
From system  (\ref{bog4}) we obtain
\begin{equation}\label{bog5}
\beta_\lambda = -p \lambda+
\sqrt{\bigg(\frac{\lambda^2}{2m}+\wt{V}_\lambda\bigg)^2-\wt{V}_\lambda^2}.
\end{equation}

In this example, we have  $u=e^{\frac ih s(x,t)}$ and
$u^*=e^{-\frac {s(x,t)}{h}}$, where $s(x,t,)= px+\beta t$, and
the variation of the action for the vector $\big({\delta u},
{\delta u^*}\big)$ is equal to $\lambda x\pm\Omega t$.

A more thorough passage to the limit gives $\wt{V}_\lambda \to V_0
=L^{-3} \int dx V(x)$.

In the classical limit, we thus obtain the famous Bogoliubov
relation (\ref{bog5}). In this case, we have $u(x)=0$, and the
exact solution coincides with the classical one as in the linear
Schr\"{o}dinger equation. The situation with  $u(x)\neq 0$  was
investigated in \cite{QuasiPart4}, and it turns out that a
relation similar to (\ref{bog5}) is the classical limit as $h\to
0$ for the variational equation in this general case. The curve
for the dependence of $\beta_\lambda$ on $\lambda$ is called the
Landau curve, and it specifies the superfluid state. The value
$\lambda_{\text{cr}}$ at which the superfluidity disappears is
called the Landau criterion. Bogoliubov explains the
superfluidity phenomenon as follows:  ``the `degenerate
condensate' can move without friction relative to elementary
excitations with an arbitrary sufficiently small velocity''
(\cite{Bogol}, p.210).

But this mathematical consideration is not related to the
Bose-Einstein condensate; merely the quasi-particle spectrum
determined for $\lambda <\lambda_{\text{cr}}$  is positive. This
means that it is metastable (see \cite{Masl_Shved}). The
Bose-Einstein condensate is mentioned here only to disprove the
idea that it follows from what was said above that this
consideration applies to a classical liquid.

Indeed, for example, the molecules of a classical undischarged
liquid are Bose particles if the number of neutrons in the
molecule is even. Because every particle (molecule) is neutral
and is formed of an even number $l$ of neutrons, an $N$-particle
equation can be written for this liquid. Thus, every its particle
is $3(2k+l)$-dimensional, where $k$ is the number of electrons;
there is a dependence on the potential $u(x_i), \ x_i \in
R^{6k+3l}$; and an equation for $N$ particles $x_i, i=1, \dots, N$
with a pair interaction potential $V(x_i-x_j)$ can be considered.

But Bogoliubov found only one series for the spectrum of the many-particle problem.
As Landau wrote, ''N.~N.~Bogoliubov recently managed to find
the general form of the energy spectrum for the Bose--Einstein gas
with a weak interaction between the particles using a clever application of second
quantization'' (\cite{landau}, p.~43).
But this series is not unique, i.e., it does not exhaust the whole energy spectrum.

In 2001, we suggested the ultrasecond quantization method
\cite{Book_Ultravt} (also see \cite{FAN_2000, Uspehi_2000,
RJ_2001, RJ_2001_2, Mas1}). The ultrasecond-quantized
Schr\"{o}dinger equations, like the second-quantized ones,
represent the $N$-particle Schr\"{o}dinger equation, and this
means that the ultrasecond-quantized equation is essentially
identical to the original $N$-particle equation: it coincides
with the latter on the $3N$-dimensional space. But in contrast to
the second-quantized case, replacing the creation and annihilation
operators with $c$-numbers does not yet give the correct
asymptotic representation; it turns out that its results coincide
with those obtained by applying the Schr\"{o}der variational
principle or the Bogoliubov variational method.

For the Bardeen exotic potential, the correct asymptotic solution
coincides with the one resulting from applying the abovementioned
ultrasecond quantization method. For potentials of general form,
in the case of pair interaction for example, the answer turns out
different. In particular, the ultrasecond quantization method
gives some other asymptotic series of eigenvalues corresponding
to the $N$-particle Schr\"{o}dinger equation, which, in contrast
to Bogoliubov series (\ref{bog5}), are not metastable. They
correspond to vortex filaments \cite{Lifshits}.

It turns out that the decisive factor here is not the
Bose-Einstein condensate but the thickness of the capillary
(nanotube) in which the liquid flows. If we consider a liquid in
a capillary or nanotube, then the velocity corresponding to
metastable states is not small for a sufficiently small radius.
Consequently, the liquid flows without friction for a smaller
velocity.

The no-flow condition on the boundary of the nanotube
(absence of flow) is the Dirichlet boundary
condition or the Born--von Karman boundary condition.
It generates a standing wave that can be
interpreted as a particle-antiparticle pair:
a particle with the momentum $p$ orthogonal to the tube
wall and an antiparticle with the momentum $-p$.

In the boson case, we consider a short-range interaction
potential $V(x_i -x_j)$. This means that only interaction with
finitely many particles is possible as $N\to\infty$ ($N$ is the
number of particles). Consequently, the potential depends on $N$
as $V_N=V((x_i-x_j)N^{1/3})$. If $V(y)$ is finitely supported in
$\Omega_V$, then the number of particles captured by the support
is independent of $N$ as $N\to\infty$. As result, superfluidity
occurs for velocities less than $\min (\lambda_{\text{cr}},
\frac{h}{2mR})$, where $R$ is the nanotube radius. The upper bound
is determined by the condition that the radius of action of the
molecule must be less than the radius of the nanotube.

We now present our own considerations that do not relate to the
mathematical presentation. Viscosity is connected with collisions
of particles: the higher the temperature is, the greater the
number of collisions. In a nanotube, there are few collisions
because only those with the tube walls occur, which is taken into
account by the series obtained below. Precisely this fact rather
than the presence of the Bose condensate leads to the weakening
of viscosity and consequently to superfluidity. In other words,
even for liquid He4, the main factor in the superfluidity
phenomenon is not the condensate but the presence of a thin
capillary  \cite{TMF_2005,RJ_2005}.

\textbf{2.} In this part of the paper we refine solutions of
equations in variations (26)-(28) and (35)-(37) presented in the
paper \cite{RJ_2007_14_4}; see also \cite{RJ_2007_14_3} and
\cite{RJ_2008_15_1}.

For the Fermi liquid (for instance, for helium 3), we solve the
mathematical problem on the reduction of the $N$-particle
Schr\"{o}dinger equation with the pair interaction potential
typical for helium: repulsion as a pair of particles approaches
and attraction as the pair of particles moves away. This means
that we do not take into account the possible resonance
interaction of Cooper pairs, as is usual in supercondactivity
problems. Moreover, our assumption forbids the radius of the
capillary to be less than the radius of the molecule itself
(otherwise the molecule simply cannot enter the capillary).
However, as the radius decreases, a superfluidity domain can
occur for a liquid with an even number of neutrons. Nevertheless,
for a Fermi liquid, superfluidity also occurs, but in
quasithermodynamics rather than in the thermodynamical limit. The
notion of quasithermodynamics was recently introduced by the
author \cite{Gol_Masl_MatZam}.

\textbf{3.} The eigenvalues of the equations in variations for
Bose-liquid are \cite{Gol_Masl_MatZam}:
\begin{equation}\label{common spectrum}
\lambda_{1,k_1k_2,l}=-\frac{\hbar^2}{m}k_1(k_2+l)+
\sqrt\frac{\xi_{k_2,l}+\sqrt{\xi^2_{k_2,l}-4\eta_{k_2,l}}}{2},
\end{equation}
where
\begin{align*}
&\xi_{k_2,l}=a^2\l \l l_{1}^2-k_2^2\r^2+\l l^2-k_2^2\r^2\r+\\&+
a\l l_{1}^2\l v_{l+3k_2}-v_{2k_2}\r+l^2\l
v_{l-k_2}-v_{2k_2}\r-k_2^2\l
v_{l-k_2}+v_{l+3k_2}-2v_{2k_2}\r\r-\\&-
(v_{l+k_2}+v_{2k_2})(v_{l+3k_2}+v_{l-k_2}-2v_{2k_2})/2,\\
&\eta_{k_2,l}=a\Big( 2a\l k_2^4-k_2^2\l
l_{1}^2+l^2\r+l_{1}^2l^2\r-\l l_{1}^2+l^2-2k_2^2\r\l
v_{l+k_2}+v_{2k_2}\r\Big)\cdot\\&\cdot\Big( 2a^2\l k_2^4-k_2^2\l
l_{1}^2+l^2\r+l_{1}^2l^2\r+\\&+ a\big(
l_{1}^2\l2v_{l-k_2}+v_{l+k_2}-v_{2k_2}\r+l^2\l
2v_{l+3k_2}+v_{l+k_2}-v_{2k_2}\r -\\&-2k_2^2\l
v_{l+3k_2}+v_{l-k_2}+v_{l+k_2}-v_{2k_2}\r\big)+\\&
+2(v_{l+3k_2}-v_{2k_2})(v_{l-k_2}-v_{2k_2})+(v_{l+k_2}+v_{2k_2})
(v_{l+3k_2}+v_{l-k_2}-2v_{2k_2})\Big)/4;
\end{align*}
and where the notation
\begin{align*}
    a=\frac{\hbar^2}{2m},\qquad l_1=l+2k_2.
\end{align*}
was used for brevity.

Relation \eqref{common spectrum}, as compared with (27) in
\cite{RJ_2007_14_3}, is uniform with respect to  $k_2$ as $k_2
\to \infty$. Note that, if $k_2 = 0$, then the Bogolyubov relation
holds,
\begin{equation}\label{Bogolub}
\lambda_{1,k_1,l}=-\frac{\hbar^2}{m}k_1l+ \sqrt{\l
\frac{\hbar^2l^2}{2m}+v_l\r^2-v_l^2}.
\end{equation}

For a system of identical Fermi-particles, the eigenvalue problem
for the system of equations in variations can similarly be
reduced to the problem of finding the eigenvalues of the equation
\begin{equation}\label{for tildelabbda fermi}
\wt\la X=MX.
\end{equation}
Here $\wt\la=\lambda+\frac{\hbar^2}{m}k_1(k_2+l)$, \ $X$ -- is a
column vector of the form
\[
X=\left(
\begin{array}{c}
u_{1,l}\\
u_{2,l}\\
v_{1,l}\\
v_{2,l}
\end{array}
\right),
\]
$M$ -- stands for the matrix
\[
M=\left(
\begin{array}{cccc}
B_1 & V & V_1 & 0\\
V & B_2 & 0 & V_2\\
M_1 & F & -B_1 & -V\\
F & M_2 & -V & -B_2
\end{array}
\right)
\]
with the elements
\begin{equation}\label{elements of M fermi}
 \begin{aligned}
    &B_1=B_{k_2,l}+\dfrac{v_{l-k_2}}{2},&&V=\dfrac{v_{l+k_2}-v_{2k_2}}{2},\\
    &B_2=B_{k_2,l+2k_2}+\dfrac{v_{l+3k_2}}{2},&&V_1=\dfrac{v_{l-k_2}-v_{l+k_2}}{2},\\
    &M_1=2i(v_{l-k_2}-v_{0})\varphi_{k_2,l},&&V_2=\dfrac{v_{l+3k_2}-v_{l+k_2}}{2},\\
    &M_2=2i(v_{0}-v_{l+3k_2})\varphi_{k_2,l+2k_2},&&F=i(v_{2k_2}-v_{l+k_2})\l\varphi_{k_2,l+2k_2}-\varphi_{k_2,l}\r,
\end{aligned}
\end{equation}
where the numbers $B_{k_2,l}$ and  $\varphi_{k_2,l}$ are
\begin{equation}\label{fi(k2,l) & b(k2,l) fermi}
\begin{aligned}
&B_{k_2,l}=\frac{\hbar^2}{2m}(l^2-k_2^2)+i(v_{l+k_2}-v_{l-k_2})\varphi_{k_2,l}-\frac{v_{2k_2}}{2},\\
&\varphi_{k_2,l}=-\frac{ib_{k_2,l}}{2}\pm\frac{i\sigma_l}{2}\sqrt{b_{k_2,l}^2-1},
\qquad
b_{k_2,l}\equiv\frac{\dfrac{\hbar^2}{m}(l^2-k_2^2)+(v_{0}-v_{2k_2})}{v_{l-k_2}-v_{l+k_2}},
\end{aligned}
\end{equation}
$k_1$, $k_2$ and $l$ are three-dimensional vectors of the form
\begin{equation}\label{Vector quasidiscrete}
2\pi\l\frac{n_1}{L_1},\frac{n_2}{L_2},\frac{n_3}{L_2}\r,
\end{equation}
$n_1$, $n_2$ and $n_3$ -- n3 are integers. The summation is taken
over all values of  $n_1$, $n_2$, $n_3$, and
\begin{align*}
&\wt{\xi}_{k_2,l}=a^2\l \l l_{1}^2-k_2^2\r^2+\l l^2-k_2^2\r^2\r
+\\&+ a\l l_{1}^2\l v_{l+3k_2}-v_{2k_2}\r+l^2\l
v_{l-k_2}-v_{2k_2}\r-k_2^2\l
v_{l-k_2}+v_{l+3k_2}-2v_{2k_2}\r\r+\\&+
(v_{l+k_2}-v_{2k_2})(v_{l-k_2}+v_{l+3k_2}-2v_{2k_2})/2,\\
&\wt{\eta}_{k_2,l}=a\Big( 2a\l k_2^4-k_2^2\l
l_{1}^2+l^2\r+l_{1}^2l^2\r+\l l_{1}^2+l^2-2k_2^2\r\l
v_{l+k_2}-v_{2k_2}\r\Big)\cdot\\&\cdot\Big( 2a^2\l k_2^4-k_2^2\l
l_{1}^2+l^2\r+l_{1}^2l^2\r+\\&+ a\big(
l_{1}^2\l2v_{l-k_2}-v_{2k_2}-v_{l+k_2}\r+l^2\l
2v_{l+3k_2}-v_{2k_2}-v_{l+k_2}\r -\\&-2k_2^2\l
v_{l+3k_2}+v_{l-k_2}-v_{l+k_2}-v_{2k_2}\r\big)+\\&
+2(v_{l+3k_2}-v_{2k_2})(v_{l-k_2}-v_{2k_2})-(v_{l+k_2}-v_{2k_2})
(v_{l+3k_2}+v_{l-k_2}-2v_{2k_2})\Big)/4,
\end{align*}
where the notation
\begin{align*}
    a=\frac{\hbar^2}{2m},\qquad l_1=l+2k_2.
\end{align*}
is used for brevity. The eigenvalues of the system of equations
in variations are
\begin{equation}\label{common spectrum fermi}
\lambda_{1,k_1k_2,l}=-\frac{\hbar^2}{m}k_1(k_2+l)+
\sqrt\frac{\wt{\xi}_{k_2,l}+\sqrt{\wt{\xi}^2_{k_2,l}-4\wt{\eta}_{k_2,l}}}{2}.
\end{equation}
For $k_2 = 0$, i.e., along a capillary, the superfluidity in the
quasithermodynamics holds and, for $k_2\neq 0$, i.e., if there is
a reflection from the walls of the capillary, then an instability
occurs because the eigenvalues become complex by \eqref{common
spectrum fermi}. This gives the same upper bound for the critical
speed as that in the Bose-case (due to occurrence of a vortex).

According to \cite{RJ_2007_14_3, RJ_2007_14_4, RJ_2008_15_1} only
the thermodynamical and quasithermo-dynamical limits in
statistical physics exist. The above solution on the
superfluidity of a Fermi liquid enables us to claim that the
phase transition in this problem from the superfluid state to the
normal one is an example of a phase transition in
quasithermodynamics.

The results obtained in this paper are based on the author's
paper of the year 1995  \cite{QuasiPart4}  which is given in
appendix below.

\newpage

\setcounter{equation}{0}

\appendix
\section*{Quasi-Particles Associated with Lagrangian Manifolds
Corresponding to Semiclassical Self-Consistent Fields. III
\footnote{V.P.Maslov, Russian J. Math. Phys. 1995, v.3, N3,
401-406.}}

\renewcommand{\theequation}{A.\arabic{equation}}

In the preceding part of this paper, we presented
Eqs.~\thetag{25} for quasi-particles associated with an
$n$-dimensional Lagrangian manifold and Eq.~\thetag{29} for
quasi-particles corresponding to a $(2n-1)$-dimensional manifold.
These equations were written out only in the $x$-chart, and the
quantum corrections were given without proof. In this part we
essentially use the canonical operator method to obtain
Eq.~\thetag{25} with corrections in the $x$-chart as well as in
any other chart of the canonical atlas \cite{1}. To derive the
correction in Eq.~\thetag{29}, a ``modified" $\dt$-function must
be used, and this will be done in the next part of the paper.

To obtain the result in an arbitrary canonical chart, one should
pass on to the $p$-representation with respect to some of the
coordinates in the Hartree equation. This is actually equivalent
\cite{2} to considering the Hartree-type equation

\begin{equation}\label{50}
\Big[H_0\Big(\str{2} {x}, -\str{1}{ih\frac{\pa}{\pa x}}\Big) +
\int dy \psi^*(y) H_1 \Big(\str {2}{x}, -\str
{1}{ih\frac{\pa}{\pa x}}; \str {2} {y}, -\str
{1}{ih\frac{\pa}{\pa y}}\Big)\psi(y)\Big]\psi(x) = \Om \psi(x),
\end{equation}
where $x,y\in\R^n$, $\psi\in L^2(\R^n)$ is a complex-valued
function, $h>0$, $\Om\in\R$, and the indices $1$ and $2$ specify
the ordering of the operators $x$ and $-ih\pa/\pa x$. The
function $H_1$ satisfies the condition
$H_1(x,p_x;y,p_y)=H_1(y,p_y;x,p_x)$. Equation~\eqref{50}
generalizes the ordinary Hartree equation (Eq.(1) in \cite{4},
where $N=1$). The study of Eq.~\eqref{50} is important, for
example, if one makes an attempt to find a solution to the
Hartree equation (1) in the momentum representation,
$$
\psi(x)=\int \wt\psi(p) e^{(i/\hbar)px}
\frac{dp}{(2\pi\hbar)^{n/2}}.
$$

Let us also discuss the variational system associated with
Eq.~\eqref{50}, which can be obtained as follows. Along with
Eq.~\eqref{50}, let us write out the conjugate equation and
consider the variations of both equations {\it assuming that the
variations\/} $\dt\psi=F$ {\it and\/} $\dt\psi^*=G$ {\it are
independent\/}.

The variational system has the form
\begin{align}\label{51}
&\Big[H_0\Big(\str {2}{x}, -\str {1}{ih\frac{\pa}{\pa x}}\Big)-\Om
 + \int dy \psi^*(y) H_1
\Big(\str {2}{x}, -\str{1}{ih\frac{\pa}{\pa x}}; \str{2} {y},
-\str{1}{ih\frac{\pa}{\pa
y}}\Big)\psi(y)\Big]F(x) \nn \\
&\quad +\int dy\Big( G(y) H_1\Big(\str{2}{x}, -\str
{1}{ih\frac{\pa}{\pa x}}; \str{2}{y}, -\str{1}{ih\frac{\pa}{\pa
y}}\Big)\psi(y) \nn \\
&\quad+\psi^*(y) H_1\Big(\str{2}{x}, -\str{1}{ih\frac{\pa}{\pa
x}}; \str{2}{y}, -\str{1}{ih\frac{\pa}{\pa
y}}\Big)F(y)\Big) \psi(x) = -\be F(x),\\
&\Big[H_0\Big(\str{1}{x}, \str{2}{ih\frac{\pa}{\pa x}} \Big)-\Om
+ \int dy \psi(y) H_1 \Big(\str{1}{x}, \str{2}{ih\frac{\pa}{\pa
x}}; \str{1}{y}, \str{2}{ih\frac{\pa}{\pa
y}}\Big)\psi^*(y)\Big]G(x)\nn \\
&\quad +\int dy\Big( F(y) H_1\Big(\str{1}{x}, \str
{2}{ih\frac{\pa}{\pa x}}; \str{1} {y}, \str{2}{ih\frac{\pa}{\pa
y}}\Big)\psi^*(y) \nn \\
&\quad +\psi(y) H_1\Big(\str{1}{x}, \str{2}{ih\frac{\pa}{\pa x}};
\str{1}{y}, \str{2}{ih\frac{\pa}{\pa y}}\Big)G(y)\Big) \psi^*(x) =
\be G(x).\nn
\end{align}
Equations~\eqref{50} and \eqref{51} play an important role in the
problem of constructing  asymptotic solutions to the $N$-particle
Schr\"odinger equation as $N\to\infty$ \cite{5}--\cite{7}.

For example, the spectrum of system \eqref{51} (possible values of
$\be$) corresponds to the spectrum of quasi-particles. Namely,
the difference between the energy of an excited state and the
ground state energy is given by the  expression
$
\sum_k \be_k n_k
$, 
where the numbers $n_k\in Z_+$, $k=\ov{1,\infty}$, which are
equal to zero starting from some $k$, define the eigenfunction
and the eigenvalue of the excited state, and $\be_k\in\R$ are the
eigenvalues of  system \eqref{51}.

In this paper we are interested in asymptotic solutions to
Eqs.~\eqref{50} and \eqref{51} as the ``inner" $h$ tends to zero.

Asymptotic solutions to Eq.~\eqref{50} are given \cite{8} by the
canonical operator on a Lagrangian manifold $\La^n=\{x=X(\al),
p=P(\al)\}$ invariant with respect to the Hamiltonian system
\begin{equation}\label{52}
\str{.}{x}=\frac{\pa H(x,p)}{\pa p}, \quad \str{.}{p}=-\frac{\pa
H(x,p)}{\pa x},
\end{equation}
where
$$
H(x,p)=H_0(x,p) + \int d\mu_\al H_1(x,p; X(\al),P(\al)),
$$
$\al \in\La^n$, and $d\mu_\al$ is an invariant measure on $\La^n$.
The Lagrangian manifold lies on the surface $H(x,p)=\Om$. If a
chart $A$ is projected diffeomorphically in the $x$-plane, then
the canonical operator acts as the multiplication by
$\exp\{(i/h)S(x)\}/\sqrt J$, where $S(x)=\int p\,dx$ on $\La^n$
and $J=Dx/D\mu_\al$. We are interested in finding asymptotic
solutions to Eqs.~\eqref{51}. Without loss of generality, we can
confine ourselves to the case of $x$-chart. Indeed, to obtain
similar expressions in the $p$-chart, one must consider the
Fourier transformation of Eqs.~\eqref{50} and \eqref{51} and
apply the same technique, since the form of the equations remains
unchanged.

Let us seek the asymptotic solutions to Eqs. \eqref{51} in the
$x$-chart in the form
\begin{equation}
F(x) = \wt f(x)\psi(x),\quad G(x)=\wt g(x)\psi^*(x), \label{53}
\end{equation}
where the functions $f$ and $g$, in contrast to $\psi$ and
$\psi^*$, have a limit as $h\to0$. One can consider a more
general case, by allowing $f$ and $g$ to be functions of $x$ and
$-ih\pa/\pa x$, but in the leading term as $h\to 0$ we have
$$
-ih\frac{\pa}{\pa x} e^{(i/h)S}\approx \frac{\pa S}{\pa
x}e^{(i/h)S},
$$
and so we arrive at functions $f$ and $g$ that depend only on $x$.

The second equation in system \eqref{51} can be rewritten in the
form
\begin{equation}
\aligned &\Big[H_0\Big(x, {ih\frac{\pa}{\pa x}} \Big) + \int dy
\psi(y) H_1 \Big( x, {ih\frac{\pa}{\pa x}};
 y, {ih\frac{\pa}{\pa
y}}\Big)\psi^*(y);\wt g(x)\Big]\psi^*(x)\\
&\quad +\int dy\Big\{ \psi(y)\wt c(y) H_1\Big(x, ih\frac{\pa}{\pa
x}; y, ih\frac{\pa}{\pa
y}\Big)\psi^*(y)\\
&\quad+\psi(y) \Big[H_1\Big( x, {ih\frac{\pa}{\pa x}};  y,
{ih\frac{\pa}{\pa y}}\Big);\wt g(y)\Big] \psi^*(y)\Big\}\psi^*(x)
= \be\wt g(x) \psi^*(x),
\endaligned
\label{54}
\end{equation}
where $[A;B]=AB-BA$ and
\begin{equation}
c(x)=\wt f(x)+\wt g(x). \label{55}
\end{equation}
Equation~\eqref{50} is used in the derivation of Eq.~\eqref{54}.
We observe that all terms containing the function $\wt g$ on the
left-hand side in Eq.~\eqref{50} are $O(h)$, since the commutator
of two operators depending on $x$ and $-ih\pa/\pa x$ is  equal,
in the classical limit, to $(-ih)$ times the Poisson bracket of
the corresponding classical quantities.

Thus, the function $c$, as well as the eigenvalue $\be$, is
assumed to be $O(h)$. Let us rescale these quantities as follows:
\begin{equation}
c(x)=h\wt c(x),\quad\be=h\wt\be. \label{56}
\end{equation}

Now we can derive the equation for $\wt g, \wt c$, and $\wt\be$
in the leading term in $h$ and the first correction to it from
Eq.~\eqref{54}, making use of the following relations:
\begin{equation}
\text{i)}\,\, \Big[A\Big(\str{1} {x}, \str{2}{ih\frac{\pa}{\pa
x}}\Big);\xi(x)\Big] =\sum_{a=1}^n i h \frac{\pa A}{\pa p_a}
\Big(\str{1} {x}, \str{2}{ih\frac{\pa}{\pa
x}}\Big)\frac{\pa\xi}{\pa x_a} - \sum_{a,b=1}^n
\frac{h^2}{2}\frac{\pa^2A}{\pa p_a \pa p_b} \Big(\str{1}{x}, \str
{2}{ih\frac{\pa}{\pa x}}\Big) \frac{\pa^2\xi}{\pa x_a \pa x_b},
\label{57}
\end{equation} where
$p_a=ih\pa/\pa x_a$, $A(x,p)$ is a function $\R^{2n}\to C$,
$\xi:\R^n\to\R$;

\hskip-0.2cm ii) $\psi(x)=\chi(x,h)e^{(i/h)S(x)}$, where
$\chi=1/\sqrt J$ in the leading term in $h$;
$$
\text{\hskip-0.1cm iii)}\,\,\qquad\qquad\qquad\qquad\qquad
ih\frac{\pa}{\pa x}e^{-(i/h)S(x)} = e^{-(i/h)S(x)} \Big(\frac{\pa
S}{\pa x}+ih \frac{\pa}{\pa x}\Big);\qquad\qquad\qquad\quad\,\,
\hfill
$$
\begin{align} \label{58}
\text{\hskip-0.1cm iv)}\qquad\quad\quad\,\,
&B\Big(ih\frac{\pa}{\pa x} + \frac{\pa S}{\pa x}\Big) = B
\Big(\frac{\pa S}{\pa x}\Big) + ih \sum_{a=1}^n \frac{\pa B}{\pa
p_a} \frac{\pa}{\pa x_a} + \frac{ih}{2}\sum_{a,b=1}^n \frac{\pa^2
B}{\pa p_a\pa
p_b} \frac{\pa^2 S}{\pa x_a\pa x_b}\nn \\
&\quad+\frac{(ih)^2}{2}\sum_{a,b=1}^n \frac{\pa^2 B}{\pa p_a\pa
p_b} \frac{\pa^2}{\pa x_a\pa x_b} +
\frac{(ih)^2}{2}\sum_{a,b,c=1}^n \frac{\pa^3 B}{\pa p_a\pa
p_b \pa p_c} \frac{\pa^2 S}{\pa x_a\pa x_b}\frac{\pa}{\pa x_c}\\
&\quad+\frac{(ih)^2}{6}\sum_{a,b,c=1}^n \frac{\pa^3 B}{\pa p_a\pa
p_b \pa p_c} \frac{\pa^3 S}{\pa x_a\pa x_b\pa x_c}\nn \\
&\quad +\frac{(ih)^2}{8}\sum_{a,b,c,d=1}^n \frac{\pa^4 B}{\pa
p_a\pa p_b \pa p_c \pa p_d} \frac{\pa^2 S}{\pa x_a\pa x_b}
\frac{\pa^2 S}{\pa x_c\pa x_d} +O(h^3), \nn
\end{align}

where all derivatives of $B$ are evaluated at the point $p=\pa
S/\pa x$.

These relations can  easily be obtained for monomial functions
$A$ and $B$. An application of formulas i)--iv) yields the
equation
\begin{align}
&i\sum_{a=1}^n \frac{\pa H}{\pa p_a^x}\frac{\pa \wt g}{\pa
x_a}(X(\al)) -
\wt\be\wt g (X(\al)) + \int d\mu_\be \wt c(X(\be))H_1\nn \\
&\quad + i\int d\mu_\be \sum_{a=1}^n \frac{\pa \wt g}{\pa
x_a}(X(\be)) \frac{\pa H_1}{\pa p_a^y}
 +\frac{h}{2} \sum_{a,b=1}^n \frac{\pa \wt g}{\pa x_a}(X(\al))
\frac{\pa^2 H}{\pa p_a^x \pa p_b^x}\frac{\pa\op{ln}J}{\pa x_b}(X(\al))\nn \\
&\quad -\frac{h}{2} \sum_{a,b,c=1}^n \frac{\pa \wt g}{\pa
x_a}(X(\al)) \frac{\pa^3 H}{\pa p_a^x \pa p_b^x \pa
p_c^x}\frac{\pa^2S}{\pa x_b \pa x_c}(X(\al))
 -\frac{h}{2} \sum_{a,b=1}^n \frac{\pa^2 \wt g}{\pa x_a \pa
x_b}(X(\al)) \frac{\pa^2 H}{\pa p_a^x \pa p_b^x} \nn \\
&\quad + \frac{ih}{2} \int d\mu_\be \wt c(X(\be)) \sum_{a,b=1}^n
\Big[\frac{\pa^2 H_1}{\pa p_a^x \pa p_b^x} \frac{\pa^2 S}{\pa x_a
\pa x_b}(X(\al)) + \frac{\pa^2 H_1}{\pa p_a^y \pa p_b^y}
\frac{\pa^2 S}{\pa x_a \pa x_b}(X(\be))\Big]\nn \\
&\quad - \frac{ih}{2} \int d\mu_\be \wt c(X(\be)) \sum_{a=1}^n
\Big[\frac{\pa H_1}{\pa p_a^x} \frac{\pa\op{ln}J}{\pa x_a}(X(\al))
+ \frac{\pa H_1}{\pa p_a^y}
\frac{\pa\op{ln}J}{\pa x_a}(X(\be))\Big]\nn \\
&\quad + \frac{h}{2} \int d\mu_\be  \sum_{a=1}^n \frac{\pa\wt
g}{\pa x_a}(X(\be)) \Big\{\sum_{b=1}^n\Big(\frac{\pa^2 H_1}{\pa
p_a^y \pa p_b^y} \frac{\pa\op{ln}J}{\pa x_b}(X(\be)) +
\frac{\pa^2 H_1}{\pa p_a^y \pa p_b^x}
\frac{\pa\op{ln}J}{\pa x_b}(X(\al))\Big)\nn \\
&\quad - \sum_{b,c=1}^n \Big(\frac{\pa^3 H_1}{\pa p_a^y \pa p_b^x
\pa p_c^x} \frac{\pa^2 S}{\pa x_b \pa x_c}(X(\al)) + \frac{\pa^3
H_1}{\pa p_a^y \pa p_b^y \pa p_c^y}
\frac{\pa^2 S}{\pa y_b \pa y_c}(X(\be))\Big)\Big\}\nn \\
&\quad -\frac{h}{2} \int d\mu_\be \sum_{a,b=1}^n \frac{\pa^2
H_1}{\pa p_a^y \pa p_b^y}\frac{\pa^2 g}{\pa x_a \pa x_b}(X(\be))
=0; \label{59}
\end{align}

in this formula the arguments
\begin{equation}
x=X(\al),\quad p^x=P(\al),\quad y=X(\be),\quad p^y=P(\be)
\label{60}
\end{equation}
of the function $H_1$ and of its derivatives, as well as the
arguments $x=X(\al)$, $p^x=P(\al)$ of the function $H$, are
omitted.

Let us now find another equation relating $\wt g$ to $\wt c$. To
this end, let us multiply the first equation in system \eqref{51}
by $\psi^*(x)$ and the second equation by $\psi(x)$. Let us
subtract  the first product from the second. We obtain
\begin{equation}\label{61}
\begin{aligned}
 &\be\psi^*(x) \psi(x)c(x) = \psi(x)
\Big[H\Big(x,ih\frac{\pa}{\pa x}\Big); \wt g(x)\Big]\psi^*(x)\\
&\quad+  \psi^*(x) \Big[H\Big(x,-ih\frac{\pa}{\pa x}\Big); \wt
g(x)\Big]\psi(x) -
 \psi^*(x)
\Big[H\Big(x,-ih\frac{\pa}{\pa x}\Big); \wt c(x)\Big]\psi(x)\\
&\quad + \psi(x)\int dy \psi(y) \Big[H_1\Big(x,ih\frac{\pa}{\pa
x}; y, ih\frac{\pa}{\pa y}\Big); \wt
g(y)\Big]\psi^*(y)\psi^*(x) \\
&\quad - \psi^*(x)\int dy \psi^*(y) \Big[\wt g(y);
H_1\Big(x,-ih\frac{\pa}{\pa x}; y, -ih\frac{\pa}{\pa
y}\Big)\Big]\psi(y)\psi(x)\\
&\quad + \psi(x)\int dy \psi(y) c(y)
H_1\Big(x,ih\frac{\pa}{\pa x}; y, ih\frac{\pa}{\pa y}\Big)\psi^*(y)\psi^*(x) \\
&\quad - \psi^*(x)\int dy \psi^*(y) H_1\Big(x,-ih\frac{\pa}{\pa
x}; y, -ih\frac{\pa}{\pa y}\Big) c(y)\psi(y)\psi(x).
\end{aligned}
\end{equation}

Let us use Eqs. \eqref{57}--\eqref{59}. We find the following
equation for $\wt g$ and $\wt c$ modulo $O(h^2)$:

\begin{align}
&i\sum_{a=1}^n \frac{\pa H}{\pa p_a^x}\frac{\pa \wt c}{\pa
x_a}(X(\al)) - \wt\be\wt c (X(\al)) - \sum_{a,b=1}^n \frac{\pa^2
H}{\pa p_a^x \pa p_b^x}
\frac{\pa^2 \wt g}{\pa x_a \pa x_b}(X(\al)) \nn \\
&\quad + \sum_{a,b=1}^n \frac{\pa^2 H}{\pa p_a^x \pa
p_b^x}\frac{\pa \wt g}{\pa
x_a}(X(\al)) \frac{\pa\op{ln}J}{\pa x_b}(X(\al))\nn \\
&\quad - \sum_{a,b,c=1}^n \frac{\pa \wt g}{\pa x_a}(X(\al))
\frac{\pa^3 H}{\pa p_a^x \pa p_b^x \pa p_c^x}\frac{\pa^2S}{\pa
x_b \pa x_c}(X(\al)) - \int d\mu_\be \sum_{a,b=1}^n \frac{\pa^2
\wt g}{\pa
x_a \pa x_b}(X(\be)) \frac{\pa^2 H_1}{\pa p_a^y \pa p_b^y}\nn \\
&\quad - \int d\mu_\be \sum_{a,b,c=1}^n \frac{\pa\wt g}{\pa
x_a}(X(\be)) \Big(\frac{\pa^2 S}{\pa x_a \pa x_b}(X(\al))
\frac{\pa^3 H_1}{\pa p_a^y \pa p_b^x \pa p_c^x} + \frac{\pa^2
S}{\pa y_a \pa y_b}(X(\be))
\frac{\pa^3 H_1}{\pa p_a^y \pa p_b^y \pa p_c^y}\Big)\nn \\
&\quad +\int d\mu_\be \sum_{a,b=1}^n \frac{\pa \wt g}{\pa
x_a}(X(\be)) \Big(\frac{\pa\op{ln}J}{\pa x_b}(X(\al)) \frac{\pa^2
H_1}{\pa p_a^y \pa p_b^x} + \frac{\pa\op{ln}J}{\pa
x_b}(X(\be))\frac{\pa^2 H_1}{\pa p_a^y \pa
p_b^y}\Big)\nn \\
&\quad - i\int d\mu_\be \wt c(X(\be)) \sum_{a=1}^n
\Big(\frac{\pa\op{ln}J}{\pa x_a}(X(\al))\frac{\pa H_1}{\pa p_a^x}
+ \frac{\pa\op{ln}J}{\pa x_a}(X(\be))
\frac{\pa H_1}{\pa p_a^y}\Big)\nn \\
&\quad+ i \int d\mu_\be \sum_{a=1}^n\frac{\pa\wt c}{\pa
x_a}(X(\be))
\frac{\pa H_1}{\pa p_a^y}\nn \\
&\quad+ i\int d\mu_\be \wt c(X(\be))\sum_{a,b=1}^n
\Big(\frac{\pa^2 H_1}{\pa p_a^x \pa p_b^x}\frac{\pa^2S}{\pa x_a
\pa x_b}(X(\al)) + \frac{\pa^2 H_1}{\pa p_a^y \pa
p_b^y}\frac{\pa^2S}{\pa x_a \pa x_b}(X(\be))\Big)\nn \\
&\quad + \frac{h}{2} \sum_{a,b,c=1}^n \frac{\pa\wt c}{\pa
x_a}(X(\al)) \frac{\pa^3H}{\pa p_a^x \pa p_b^x \pa
p_c^x}\frac{\pa^2S}{\pa x_b \pa x_c}
(X(\al))\nn \\
&\quad - \frac{h}{2} \sum_{a,b=1}^n \frac{\pa\wt c}{\pa
x_a}(X(\al)) \frac{\pa^2 H}{\pa p_a^x \pa p_b^x}
\frac{\pa\op{ln}J}{\pa x_b}(X(\al)) + \frac{h}{2}\sum_{a,b=1}^n
\frac{\pa^2 H}{\pa p_a^x \pa p_b^x}\frac{\pa^2\wt c}{\pa x_a \pa
x_b}(X(\al))\nn \\
&\quad + \frac{h}{2} \int d\mu_\be \sum_{a,b,c=1}^n \frac{\pa\wt
c}{\pa x_a}(X(\be)) \Big( \frac{\pa^3H_1}{\pa p_a^y \pa p_b^x \pa
p_c^x}\frac{\pa^2S}{\pa x_b \pa x_c} (X(\al)) +
\frac{\pa^3H_1}{\pa p_a^y \pa p_b^y \pa p_c^y}\frac{\pa^2S}{\pa
x_b \pa x_c}
(X(\be))\Big)\nn \\
&\quad +  \frac{h}{2} \int d\mu_\be \sum_{a,b=1}^n
\Big[\frac{\pa^2 H_1}{\pa p_a^y \pa p_b^y}\Big( \frac{\pa^2\wt
c}{\pa x_a \pa x_b}(X(\be)) -
\frac{\pa\wt c}{\pa x_a}(X(\be)) \frac{\pa\op{ln}J}{\pa x_b}(X(\be))\Big)\nn \\
&\quad - \frac{\pa\wt c}{\pa x_a}(X(\be)) \frac{\pa\op{ln}J}{\pa
x_b}(X(\al)) \frac{\pa^2 H_1}{\pa p_a^y \pa p_b^x}\Big]=0.
\label{62}
\end{align}
If $H_0(x,p_x) = p_x^2/2 +U(x)$ and $H_1(x,p_x;y,p_y) = V(x,y)$,
then Eqs.~\eqref{59} and \eqref{62} become much  simpler and
acquire the form
\begin{equation}\label{63}
\aligned (i\nabla S\nabla -\wt\be)\wt g + \int V(x,X(\al'))\wt
c(X(\al'))\,d\mu_{\al'}
+ \frac{h}{2}(-\De\wt g+\nabla\op{ln}J\nabla \wt g)&=0,\\
(i\nabla S\nabla -\wt\be)\wt c -\De\wt g +\nabla \op{ln}J\nabla
\wt g - \frac{h}{2}(-\De\wt c+\nabla\op{ln}J\nabla \wt c)&=0.
\endaligned
\end{equation}
From Eqs.~\eqref{63} one can approximately find the functions $F$
and $G$, which are important for constructing approximate wave
functions in the $N$-particle problem as $N\to\infty$ \cite{5}.

Let us now relate the obtained results to the solution to
variational equation for the Vlasov equation, obtained in the
preceding part of this paper \cite{3}.

Let $\wh\rho$ be the projection on the function $\psi$. Its kernel
is $\wt\rho(x,y)=\psi(x)\psi^*(y)$, and its symbol is $\rho(x,p)
=\psi(x)\wt\psi^*(p) e^{(i/\hbar)px}$. The operator $\wh\rho$
satisfies the Wigner equation, which  reduces to the Vlasov
equation as $h\to0$. The operator $\wh\si$ with the kernel
$F(x)\psi^*(y)+\psi(x)G(y)$ is equal to
\begin{equation}\label{64}
\wh\si = \wt f\wh\rho +\wh\rho\,\wt g
\end{equation} and
satisfies the variational equation to the Wigner equation, which
is reduced to the variational equation for the Vlasov equation
\thetag{20} obtained in [3]. In Eq.~\eqref{64} $\wt f$ and $\wt g$ are the
operators of multiplication by the functions $\wt f$ and $\wt g$.
We see that in the semiclassical approximation the symbol of
$\si$ is $O(h)$, since $\wh\si=[\wh\rho;\wt g\,]+\hbar\,\wt c\,
\wh\rho$ and
$$
\si(x,p) \simeq \hbar\Big(-i\sum_{a=1}^n \frac{\pa\rho}{\pa
p_a}(x,p) \frac{\pa\wt g}{\pa x_a}+\wt c\rho\Big).
$$
Since $\rho$ is the $\dt_\La$-function in the semiclassical
approximation  \cite{3}, the function $\si$ is actually the sum
of the $\dt_\La$-function and its derivative. Equations~\eqref{63}
are consistent with Eqs.~\thetag{24} obtained in \cite{3} for the
coefficients of $\dt$ and $\dt'$. Thus, the approach suggested in
this part allows us to find an asymptotic formula for $\si$ as
well.

The author is deeply grateful to O.~Yu.~Shvedov, whose assistance
in carrying out all computations was invaluable.


\begin{thebibliography}{99}

\bibitem{Arxiv_1}
V.~ P.~Maslov,
On the Dispersion Law of the Form
$\varepsilon(p)={\hbar^2p^2}/{2m}+\wt{V}(p)-\wt{V}(0)$ for
Elementary Excitations of a Nonideal Fermi Gas in the Pair
Interaction Approximation  $(i\leftrightarrow j), \
V(|x_i-x_j|)$. // arXiv:0710.0537v2 [math-ph] 4 Oct 2007.

\bibitem{ArXiv_2}
V. P. Maslov, On the Superfluidity of Classical Liquid in
Nanotubes.// arXiv:0708.0919v1 [math-ph] 7 Aug 2007.

\bibitem{VKB}
V.~P.~ Maslov, Complex Method of WKB in Nonlinear Equations [in
Russian], Nauka, Moscow (1977).
English transl.:  Complex WKB Method for Nonlinear Equations:
I. Linear Theory (Progr. Phys., Vol.~16), Birkhauser, Basel (1994).

\bibitem {Bogol}
N.~N.~Bogoliubov, ``Towards a Theory of Superfluidity''
in: Selected Works in Three Volumes, Vol.~2, Naukova
Dumka, Kiev (1970), pp. 210--224 [in Russian].

\bibitem {QuasiPart4}
V.~P.~Maslov. ``Quasi-Particles Associated with Lagrangian Manifolds
Corresponding to Semiclassical Self-Consistent Fields. III,''
Russian J. Math. Phys. v.3, N3, 401--406  (1995).

\bibitem{Masl_Shved}
V.~P.~Maslov and O.~Yu.~Shvedov, ``The Complex Germ Method in
Many-Particle Problems and Problems of Quantum Field Theory,''
URSS, Moscow (2000) [in Russian].

\bibitem {landau}
L.~D.~Landau, "Toward a Theory of Superfluidity''
in:   Collected Works, Vol. 2, Nauka, Moscow (1969), pp.~42--46 [in Russian];
L.~D.~Landau, Phys. Rev., v.75, 884--885 (1949).

\bibitem{Book_Ultravt}
V.~P.~Maslov, Quantization of Thermodynamics and Ultrasecond
Quantization, Institute for Computer Studies,
Moscow (2001) [in Russian].

\bibitem{FAN_2000}
V.~P.~Maslov, Funct. Anal. Appl., v.34, 265--275 (2000).

\bibitem{Uspehi_2000}
V. P. Maslov, Russ. Math. Surveys, v.55, 1157--1158 (2000).

\bibitem{RJ_2001}
V.~P.~Maslov, ``Some Identities for Ultrasecond-Quantized Operators,''
Russian J. Math. Phys. v.8, N3, 309--321 (2001).

\bibitem{RJ_2001_2}
V.~P.~Maslov,
``Quantization of Thermodynamics, Ultrasecondary Quantization and a New Variational Principle,''
Russian J. Math. Phys. v.8, N1, 55--82 (2001).

\bibitem{Mas1}
V.~P.~Maslov, Theoret. Math. Phys., v.132, 1222--1232 (2002).

\bibitem{Lifshits}
E.~M.~Lifshits and L.~P.~Pitaevskii,
Statistical Physics: Part~2. Theory of Condensed State, Nauka, Moscow (1978)
[in Russian]; English transl., Pergamon, Oxford (1980).

\bibitem{TMF_2005}
V.~P.~Maslov, Theoret. Math. Phys., v.143, 741--759 (2005).

\bibitem{RJ_2005}
V.~P.~Maslov. ``Resonance between One-Particle (Bogoliubov) and
Two-Particle Series in a  Superfluid Liquid in a Capillary,''
Russian J. Math. Phys. v.12, N3, 369--379 (2005).

\bibitem{RJ_2007_14_3}
V.~P.~Maslov, ``On the Superfluidity of Classical Liquid in
Nanotubes, I. Case of Even Number of Neutrons,''
Russ. J. Math. Phys. v.14, N3, 304--318 (2007).

\bibitem{RJ_2007_14_4}
V.~P.~Maslov, ``On the Superfluidity of Classical Liquid in
Nanotubes,  II.  Case of Odd Number of Neutrons,''
Russ. J. Math. Phys. v.14, N4, 453--464 (2007).

\bibitem{RJ_2008_15_1}
V.~P.~Maslov,  ``On the Superfluidity of Classical Liquid in
Nanotubes,  III,''
Russ. J. Math. Phys. v.15, N1, 61--65 (2008).

\bibitem{Gol_Masl_MatZam}
D.~S.~Golikov and V.~P.~Maslov,
``On the Exact Solution of the Four-Row Matrix Corresponding
to the Variational Equations for Ultrasecond Quantization Problems,''
Mat. Zametki v.83, N2, 305--309 (2008).

\end{thebibliography}

\begin{thebibliography}{99}
\bibitem{1}
Maslov, V. P., Th\'eorie des Perturbations et M\'ethodes
Asymptotiques, Dunod, Paris, 1972.

\bibitem{2}
Maslov, V. P., ``Equations of Self-Consistent Field,''
in Sovremennye Problemy Matematiki, vol.~11, 1978, pp.~153--234.

\bibitem{3}
Maslov, V. P., ``Quasi-Particles Associated with Lagrangian
Manifolds Corresponding to Classical Self-Consistent Fields. II,''
Russian J. of Math. Phys. v.3,  N1, 123--132 (1995).

\bibitem{4}
Maslov, V. P., ``Quasi-Particles Associated with Lagrangian
Manifolds Corresponding to Classical Self-Con\-sistent Fields. I,''
Russian J. of Math. Phys. v.2, N4, 528--534 (1994).

\bibitem{5}
Maslov, V. P., and Shvedov, O. Yu.,
``Quantization in the Neighborhood of Classical Solution
in the $N$-Particle Problem and Superfluidity,''
Theoret. and  Math. Phys. v.98, N2, 181--196 (1994).

\bibitem{6}
Maslov, V. P., and Shvedov, O. Yu.,
``Complex WKB-Method in the Fock Space,''
Dokl. Akad. Nauk, v.340, N1, 42--47 (1995).

\bibitem{7}
Maslov, V. P., and Shvedov, O. Yu.,
``Large deviations in the many-body problem,''
Mat. Zametki v.57, N1, 133--137 (1995).

\bibitem{8}
Maslov, V. P.,
Complex Markov Chains and Feynman Path Integral,  Nauka, Moscow, 1976.

\end{thebibliography}
\end{document}